\documentclass{llncs}%
\usepackage{makeidx}
\usepackage{graphicx}
\usepackage{url}
\usepackage{amsmath,amssymb}
\usepackage{amsmath}
\usepackage{amsfonts}
\usepackage{amssymb}%
\begin{document}
\title{Simulation of the 2009 Harmanli fire (Bulgaria)}
\author{Georgi Jordanov\inst{1} Jonathan D. Beezley\inst{2}
Nina Dobrinkova\inst{3}\\ Adam K. Kochanski\inst{4} Jan Mandel\inst{2}
Bed\v{r}ich Soused\'{\i}k\inst{2}
}
\institute{Institute of Geophysics, Bulgarian Academy of Sciences, Sofia\\
\email{gjordanov@geophys.bas.bg}
\and Department of Mathematical and Statistical Sciences,\\
University of Colorado Denver, Denver, CO\\
\email{\{jon.beezley.math,jan.mandel,bedrich.sousedik\}@gmail.com}
\and Institute of Information and Communication Technologies\\
Bulgarian Academy of Sciences, Sofia\\
\email{ninabox2002@gmail.com}
\and Department of Meteorology, University of Utah, Salt Lake City, UT\\
\email{adam.kochanski@utah.edu}
}
\maketitle\begin{abstract}
We use a coupled atmosphere-fire model to simulate a fire
that occurred on August 14--17, 2009, in the Harmanli region, Bulgaria.
Data was obtained from GIS and satellites imagery, and from standard
atmospheric data sources. Fuel data was classified in the 13 Anderson
categories. For correct fire behavior, the spatial resolution of the models needed
to be fine enough to resolve the essential micrometeorological effects.
The simulation results are compared to available incident data.
The code runs faster than real time on a cluster. The model is available from 
\url{openwfm.org} and it extends WRF-Fire from WRF 3.3 release.
\end{abstract}%

\section{Introduction}

Research in the southern member states of European Union (EU) in the last
30 years noted very high increase of the forest fires on their teritories and Bulgarian
statistic shows a similar trend~\cite{Dobrinkova-2011-WAB}. A team from
Bulgarian Academy of Sciences has started a literature-based analysis on the
available models in 2007 and a wildland-fire modeling initiative in 2008, and
they have selected the Weather Research and Forecasting model with fire
behaviour module, WRF-Fire \cite{Mandel-2011-CAF,Mandel-2009-DAW}. In 2009, it
was decided to select a real fire from the national database maintained by the
Ministry of Agriculture, Forest and Food, administrative division Forests and
Forest Protection, for a demonstration of the model. The objective of the present
work is to demonstrate the simulation capabilities of the model with real input data.

There were 108 forest fires in 2009, with total 18,105 hectars of forest
burned. In most of the cases, the fires have started from the surrounding area
with tall grasses and different types of bushes. 15,072 hectars of these non
forest areas burned, which showed to the authorities that without prevention
by prescribed burns, burning grass with two-three years old layers can easily
turn into a forest fire. We have chosen as the case study a large forest fire
close to town Harmanli, Bulgaria, on August 14--17, 2009. This fire is said to
be caused by a barbeque on land with tall grasses and bushes near to the
forest, which turned into a three-day forest fire. The area of the fire is on the
south border of the protected zone ``Ostar Kamak,'' a part of the network 
NATURA 2000,
near Bulgaria-Greek-Turkey border.

The aim of this paper is to describe how input data, obtained in 
Bulgaria, may be used to simulate fire behavior, and to compare the results with the
observed fire. WRF-Fire was used to simulate a wild fire in Bulgaria
\cite{Dobrinkova-2011-WAB} before with partly ideal data. This is the first
use with real data and an assessment its capability for prediction. If further
validation is satisfactory, this model might be used in forecast mode in future.

\section{Summary of the model}

The model couples the mesoscale atmospheric code
WRF-ARW~\cite{Skamarock-2008-DAR} with a fire spread module,
based on the Rothermel model \cite{Rothermel-1972-MMP} and implemented by the
level set method. It has grown out of CAWFE
\cite{Clark-1996-CAF,Clark-2004-DCA}, which couples the Clark-Hall atmospheric
model with fire spread implemented by tracers. The atmospheric model supports
refined meshes, called domains. Only the finest domain is coupled with the
fire model. See \cite{Mandel-2011-CAF,Mandel-2009-DAW} for
futher details and references.

\section{Data sources}

Collecting input data and making it usable for the model is a major component
of the work necessary to simulate fire behavior. A WRF-Fire simulation
requires input data from a variety of sources from meteorological initial and
boundary conditions to static surface properties. Because WRF is a mesoscale
meteorological model, typical data sources are only available at resolutions
ranging from $10$ to $100$ km, while our simulations occur on meshes of
resolution $1,000$ times finer. For this simulation of the Harmanli fire, we
employ the highest resolution available to us. As more accurate and higher
resolution data become available in the future, more detailed simulations will
become possible.

For the meteorological inputs, we use a global reanalysis from the U.S.
National Center for Environmental Protection (NCEP). This data is given on a
$1$ degree resolution grid covering the entire globe with $6$ hour reanalysis
cycle. The data is freely available and can be downloaded automatically over
\texttt{HTTP} using a simple script. The data is downloaded as gridded binary
(GRIB) files, which are extracted into an intermediate format using a utility
called \texttt{ungrib} included in the WRF preprocessing system (WPS).
Although the resolution of the NCEP global analysis is limited, it still may be
useful as data source of data for model initialization in multi-domain setups. 
While there are many local sources of finer
resolution meteorological data available, they must be obtained individually
for each region and combined into a single source. This process is labor
intensive and cannot be automated on demand. Because our ultimate goal is to run the
model in real time to forecast ongoing events, the use of such data is impractical.

Creating the simulation also requires a number of static data fields
describing the surface properties of the domain. All such data is available as
part of a standard global dataset for WRF. The
fields in this dataset are available at various resolutions ranging from about
$1$ km to $10$ km, which is sufficient for most mesoscale weather modeling
purposes. Each field is stored in a unique format consisting of a series of
simple binary files described by a text file. The \texttt{geogrid} utility in
WPS interpolates the data in these files onto the model grid and produces an
intermediate \texttt{NetCDF} file used in further preprocessing steps. While
the standard \texttt{geogrid} dataset is sufficient for most weather
forecasting applications, it lacks two high resolutions fields. These fields,
surface topography and fuel information, are essential for accurately modeling
fire behavior because they directly affect the rate of spread of the fire
front inside the model.

While topography data is provided to WPS from the standard source, it comes
from the USGS $30$ arc second resolution global dataset (GTOPO30), which lacks
enough detail to be used for our purposes. As a result, much more detailed
source for the area of Harmanli from the Shuttle Radar Topography Mission
(SRTM) at \url{http://eros.usgs.gov} is used, which provides topography at a
resolution of about $90$ m.

\begin{table}[pt]
\begin{center}%
\begin{tabular}
[c]{l|l}%
Category & Description\\ \hline
1 & Artificial, non-agricultural vegetated areas (141,142)\\
2 & Sport Complex,Irrigated Cropland and Pasture,Bare Ground Tundra,\\
& Arable land (211,212,213), Open spaces with little or no\\
& vegetation (331,332,333,334,335)\\
3 & Cemeteries, Dryland Cropland and Pasture, Grassland, Permanent\\
& crops(221,222,223), Pastures (231), Heterogeneous agricultural\\
& areas(241,242,243,244), Scrub and/or herbaceous vegetation\\
& associations (321,322,323,324)\\
4 & Herbaceous Tundra, Parks\\
5 & Wooded Wetland\\
6 & Wooded Tundra, Orchard\\
7 & Mixed Forest\\
8 & Deciduous Needleleaf Forest, Forests (311,312,313)\\
9--13 & N/A\\
14 & Urban fabric (111,112), Industrial, commercial, and\\
& transport units (121,122,123,124), Mine, dump and construction\\
& sites (131,132,133), Wetlands(411,412,421,422,423), Water\\
& bodies (511,512,521,522,523)
\end{tabular}
\end{center}
\caption{Fuel categories from satellite imagery and CORINE code (in parentheses).}%
\label{tab:categories}%
\end{table}

\begin{figure}[t]
\begin{center}
\includegraphics[height=3.2in]{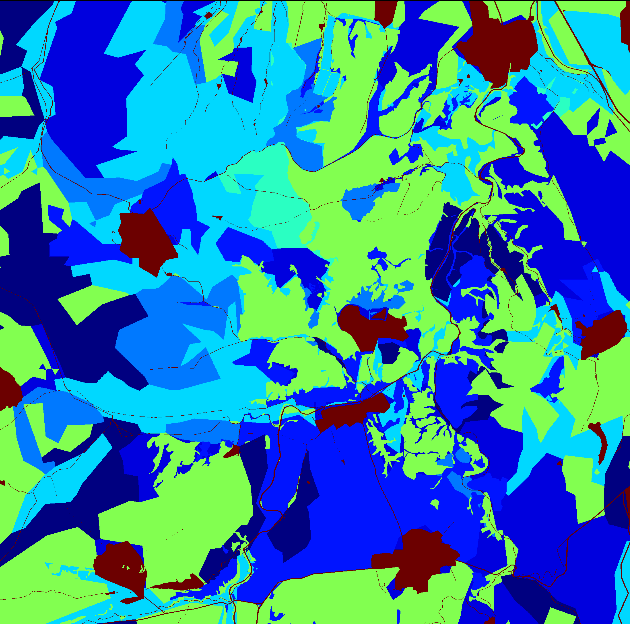}
\includegraphics[height=3.2in]{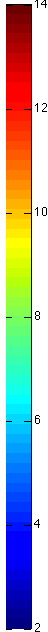}
\end{center}
\caption{Fuel map of the simulation area. The colors are the 13 Anderson fuel categories.}%
\label{fig:fuels}%
\end{figure}

\begin{figure}[t]
\begin{center}
\includegraphics[height=3.2in]{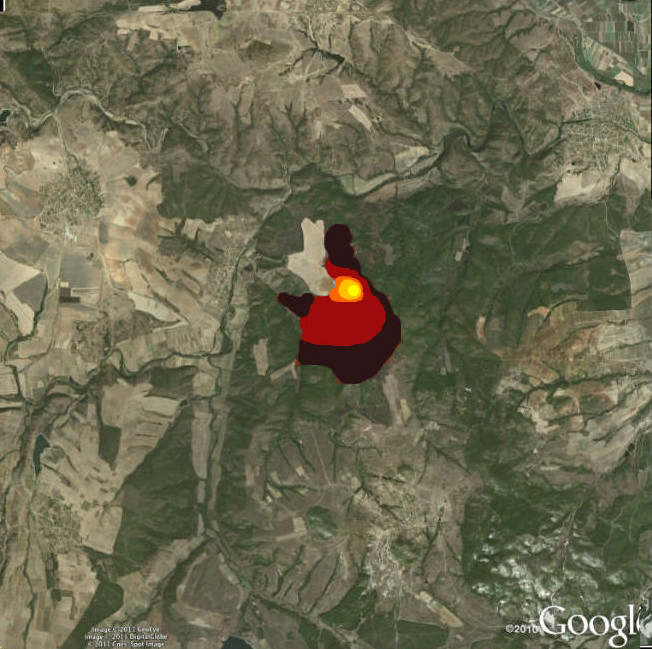}
\end{center}
\caption{Burned area every 4 hours after the fire start.}%
\label{fig:merge2}%
\end{figure}

\begin{figure}[t]
\begin{center}
\includegraphics[height=3.2in]{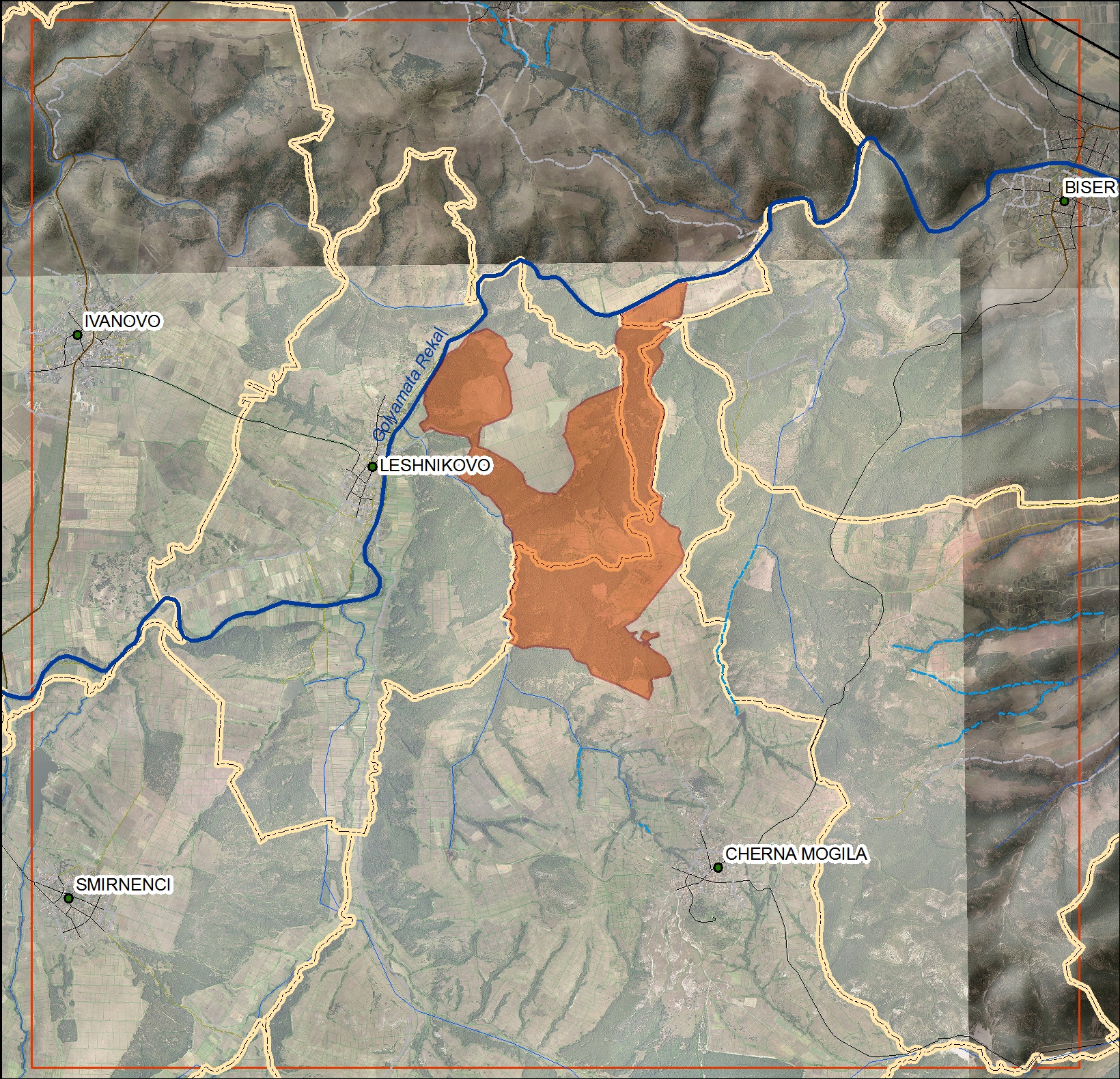}
\end{center}
\caption{Map of  the burned area and the fire perimeter.}%
\label{fig:burn_area}%
\end{figure}

\begin{figure}[t]
\begin{center}
\includegraphics[height=3.2in]{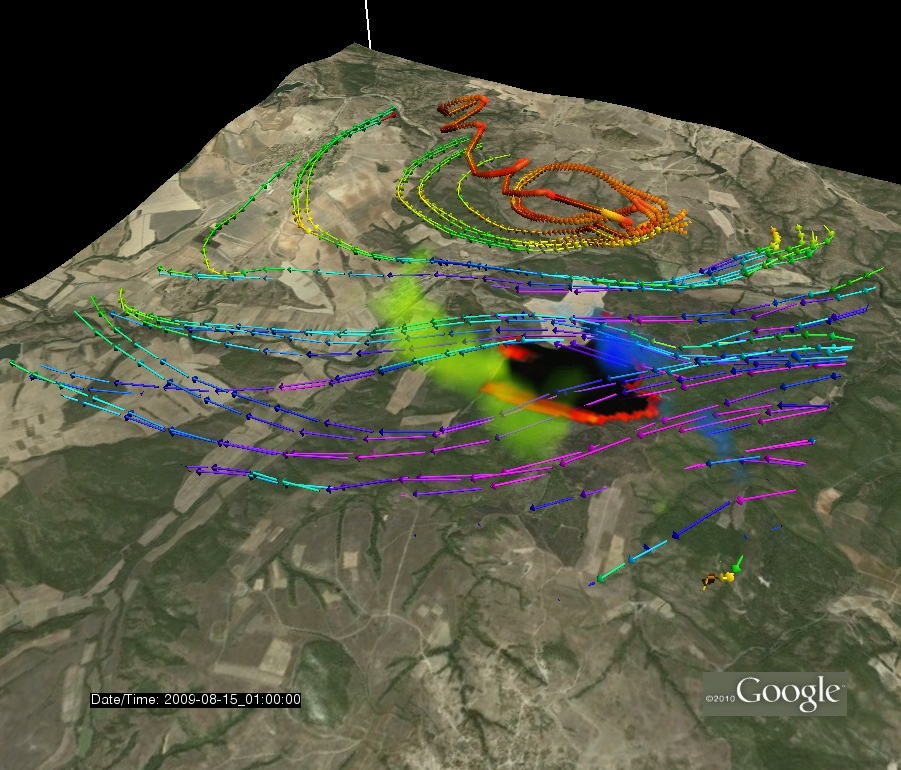}
\end{center}
\caption{The heat flux (red is high), the burned area (black), and the atmospheric flow (purple is over $10$m/s). Note the updraft caused by
the fire. Ground image from Google Earth.}%
\label{fig:flow_closeup}%
\end{figure}

The data received from the server is a GIS raster format (DTED), which must
first be processed and converted to \texttt{geogrid}'s binary data format.
Using the powerful and open-source Quantum GIS (\url{www.qgis.org}), we open
the downloaded raster and fill in the missing data with bilinear interpolation
and project the data onto the Lambert Conformal Conic projection used in the
model. Finally, we export the raster into a \texttt{GeoTIFF} file, which can
be converted into the \texttt{geogrid} format using a utility included with
the extended source distribution from \url{openwfm.org}, called \texttt{convert\_geotiff}. This
procedure is explained in detail at
\url{http://www.openwfm.org/wiki/How_to_run_WRF-Fire_with_real_data} .

The final piece of surface data needed for input into \texttt{geogrid} is a
categorical field describing the properties of the fuels. In the U.S., this
data is readily available from the USGS; however, no such data exists for the
Harmanli region. Instead, we create this field using data from the Corine
Landcover Project (financed by the European Environment Agency and the member
states). This project provides landcover data for Bulgaria with $100$ m
resolution with a $25$ ha minimum mapping unit
(\url{http://www.eea.europa.eu/data-and-maps/data/corine-land-cover-2006-raster}).
We use downloaded data along with orthophoto data from the geoportal of the
Ministry of Regional Development and Public Works (MRDPW) of Bulgaria to
estimate the fuel behavior throughout the domain. All rivers, lakes, villages
and forest areas have been vectorized using the orthophoto images combined
with CORINE2006 into a GIS vector shape file. The vectorized file provides
very high accuracy of representation for non burning areas like rivers and
lakes as well as areas with high burning fuel level like woods. We assess the
fuel behavior of every land cover category using data from the MRDPW and
assign each area a fuel category using the 13 standard Anderson fuel models
\cite{Anderson-1982-ADF}. Table~\ref{tab:categories} gives a description of
the fuel categories used in the Harmanli simulation.

This fuel level data combined with the vectorized landcover areas gives us a
final shape file with attributes for each polygon fuel level. We again use
Quantum GIS to rasterize the shape file into a $10$ m resolution
\texttt{GeoTIFF} file, which we convert into the \texttt{geogrid} format as
described above. Along with WPS's standard global datasets, we place the newly
created files in the WPS working directory and run the \texttt{geogrid}
binary. The resulting
input files contain all the standard WRF fields along with several additional
variables generated from the high resolution topography and fuel categories.

\section{Simulation results}

The atmospheric model was run with two domains. The outer one with $250$m resolution consists of 
$180\times 180\times 41$ grid points, while the inner one with $50$m resolution consists of
$221 \times 221 \times 41$ grid points. The fire model (coupled with the inner domain)
runs with mesh step $5$m. The time step for the inner domain and the fire model was $0.3$s,
while for the outer domain it was $1.5$s.

The fuel data is in Fig.~\ref{fig:fuels}. The simulated fire spread is shown in
Fig.~\ref{fig:merge2}. The actual final burn area
is shown in Fig.~\ref{fig:burn_area} for comparison.
Fig.~\ref{fig:flow_closeup} shows the atmospheric flow above the fire.

\begin{table}[t]
\begin{center}
\begin{tabular}{l|rrrrrrrrrrrr}
Cores & 6 & 12 & 24 & 36 & 60 & 120 & 240 & 360 & 480 & 720 & 960 & 1200\\ \hline
Fire & 1.91 & 1.08 & 0.50 & 0.34 & 0.22 & 0.13 & 0.08 & 0.06 & 0.06 & 0.04 & 0.10 & 0.04\\
Inner domain & 6.76 & 7.05 & 2.90 & 2.06 & 1.20 & 0.73 & 0.45 & 0.32 & 0.26 & 0.23 & 0.24 & 0.17\\
Outer domain & 0.00 & 0.00 & 0.00 & 0.02 & 0.02 & 0.04 & 0.04 & 0.06 & 0.06 & 0.08 & 0.07 & 0.15\\
Total & 10.59 & 9.21 & 3.91 & 2.75 & 1.64 & 0.99 & 0.61 & 0.44 & 0.37 & 0.31 & 0.44 & 0.26
\end{tabular}
\end{center}
\caption{Execution times divided by simulation time for increasing number of processor cores. 
The fraction is given separately for the 3 
components of the simulation: fire model and the two atmospheric model domains. 
The outer
domain time includes communication between the atmospheric domains.
}
\label{tab:parallel}
\end{table}

\begin{figure}[t]
\begin{center}
\begin{tabular}{cc}
\includegraphics[width=2.4in]{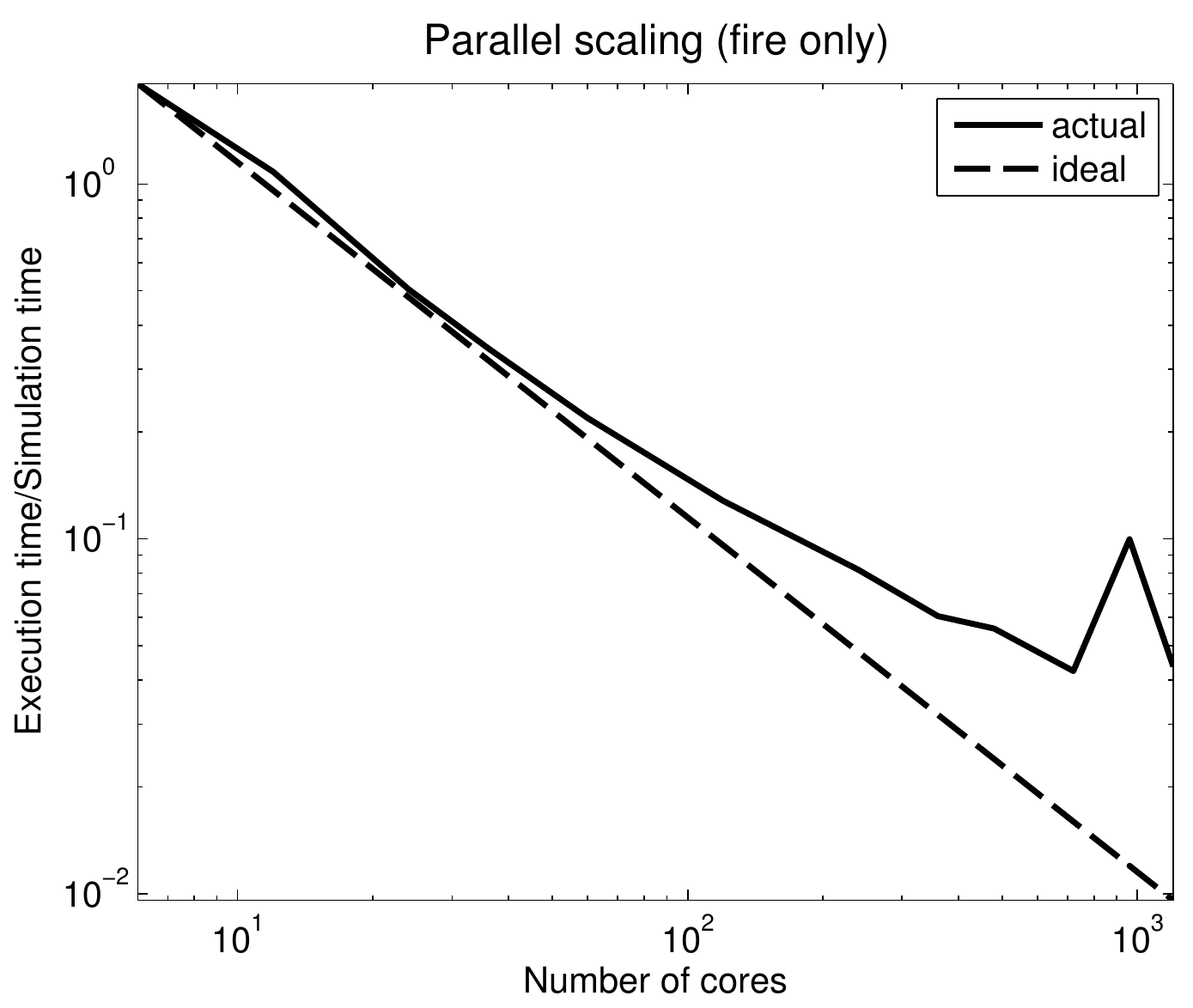} &
\includegraphics[width=2.4in]{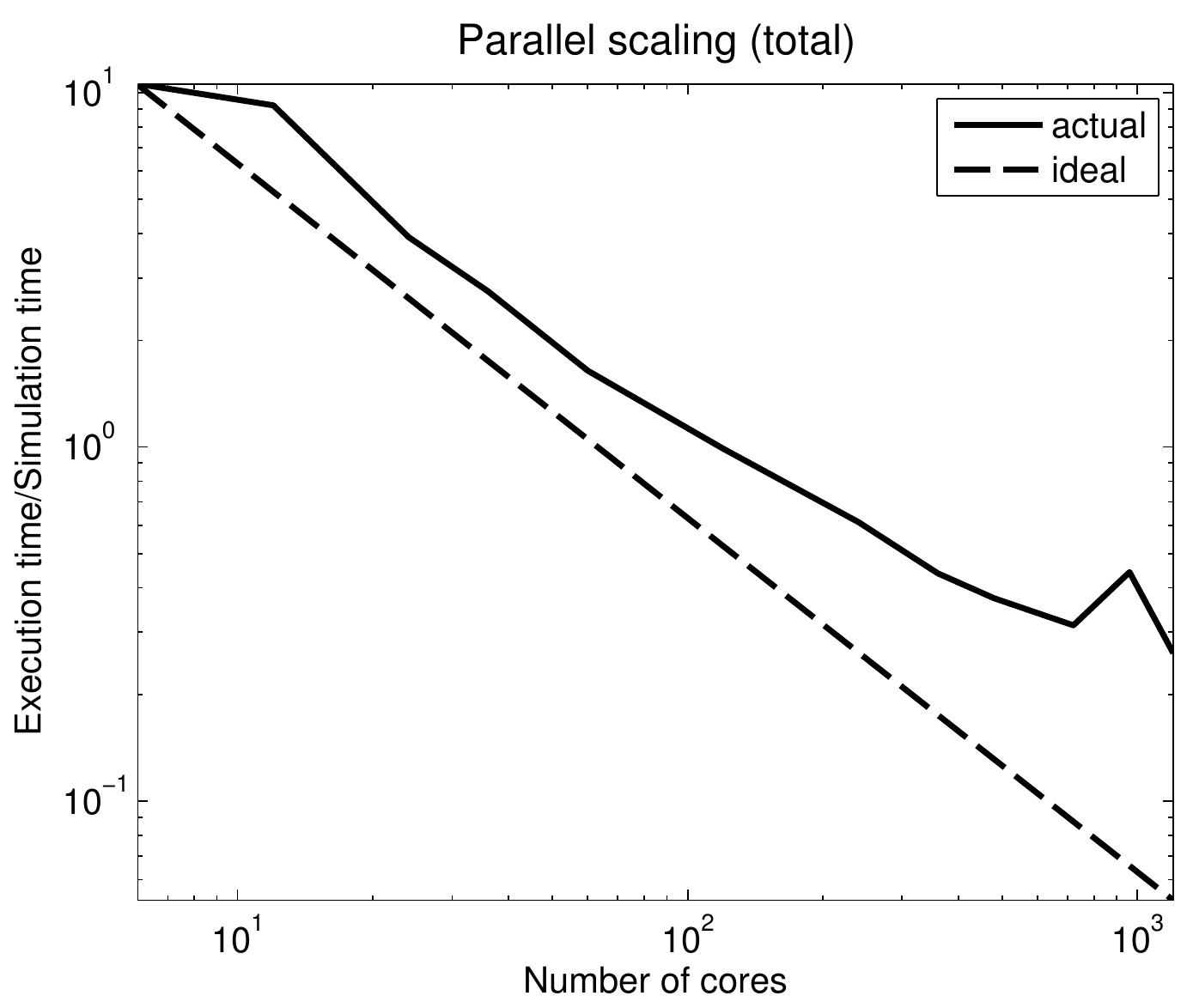}
\end{tabular}
\end{center}
\caption{Parallel scalability data plotted from Table \ref{tab:parallel}.}%
\label{fig:scalability}%
\end{figure}

\section{Parallel performance}

The speed of the simulation is essential, because only a model that is faster
than real time can be used for forecasting. We have performed computations on
the Janus cluster at the University of Colorado. The computer consists of
nodes with dual Intel X5660 processors (total 12 cores per node), connected by
QDR InfiniBand. Table ~\ref{tab:parallel} and Figure~\ref{fig:scalability} show that the coupled model is
capable of running faster than real time, and the performance scales well with
the number of processors.
The model 
runs slightly faster than real time (1s of simulation time takes 0.99s to compute) on 120 cores. 

\section{Conclusion}

We have demonstrated wildfire simulation based on real data in
Bulgraria from satellite measurement and existing GIS databases. While the
simulation provides a reasonable reproduction of the fire spread, further
refinement is needed. Data assimilation \cite{Mandel-2009-DAW,Mandel-2010-FFT}
will also play an important role. As seen from comparison of Figures~\ref{fig:merge2} and \ref{fig:burn_area},
there is a good, but not not complete, agreement of the simulated and the real burned area. Despite this, 
the simulation showed correct fire line propagation,
and it can give forecast and valuable information for future firefighting actions in different areas
with different meteorological conditions. The model can perform faster than real time
at the required resolution, thus satisfying one basic requirement for a future use for prediction.

\subsubsection*{Acknowledgements}

This work was partially supported by the National Science Fund of the Bulgarian Ministry of Education,
Youth and Science under Ideas Concourse Grant DID-02-29 ``Modelling Processes with Fixed Development Rules (ModProFix),''
ESF project number BG51PO001-3.3.04-33/28.08.2009,
U.S. National Science Foundation (NSF) under grant AGS-0835579, and by U.S. National Institute of Standards and Technology
Fire Research Grants Program grant 60NANB7D6144. Computational resources were provided by
NSF grant CNS-0821794, with additional support from UC Boulder, UC Denver, and
NSF sponsorship of NCAR.

\bibliographystyle{splncs}
\bibliography{Sozopol_2011}

\end{document}